# Macromolecular Dynamics in Red Blood Cells Investigated Using Neutron Spectroscopy

Andreas Maximilian Stadler[1,*], Lambert van Eijck[2], Franz Demmel[3], and Gerhard Artmann[4]

[1] Research Centre Jülich, 52425 Jülich, Germany

[2] Institut Laue-Langevin, 38042 Grenoble, France

[3] Rutherford Appleton Laboratory, Didcot OX11 0QX, United Kingdom

[4] Institute of Bioengineering, Aachen University of Applied Science, 52428 Jülich, Germany

* corresponding author: a.stadler@fz-juelich.de

# Abstract

We present neutron scattering measurements on the dynamics of hemoglobin (Hb) in human red blood cells *in vivo*. Global and internal Hb dynamics were measured in the *ps* to *ns* time- and Å length-scale using quasielastic neutron backscattering spectroscopy. We observed the cross-over from global Hb short-time to long-time self-diffusion. Both short- and long-time diffusion coefficients agree quantitatively with predicted values from hydrodynamic theory of non-charged hard-sphere suspensions when a bound water fraction of around 0.23g $H_2O$/ g Hb is taken into account. The higher amount of water in the cells facilitates internal protein fluctuations in the *ps* time-scale when compared to fully hydrated Hb powder. Slower internal dynamics of Hb in red blood cells in the *ns* time-range were found to be rather similar to results obtained with fully hydrated protein powders, solutions and *E. coli* cells.

# Introduction

Ongoing research is dedicated to obtaining a coherent picture of the interactions and dynamical properties of proteins in their physiological environment. Cells are highly complex objects which are composed of organelles, tens of thousands of different proteins, RNA and DNA, lipids, polysaccharides and many other chemical components. Red blood cells (RBC) in this sense are exceptional. They are highly specialized and relatively simple in their composition with the main macromolecular component hemoglobin (Hb) making of 92% of the dry weight. The concentration of Hb in RBC is c=0.33 g/ml with a corresponding volume fraction of $\phi$=0.25 (Krueger and Nossal 1988). The hydrodynamic radius of Hb is ~32 Å (Digel et al. 2006), and the average distance between Hb molecules is in the order of 1 nm (Krueger and Nossal 1988). RBC are therefore particularly well suited model systems to study the physical properties of concentrated protein solutions *in-vivo*.

From a biological point of view the properties of human RBC are interesting to study as well as they exhibit a variety of remarkable properties. RBC have been shown to undergo a passage transition through narrow micropipettes at body temperature (Artmann et al. 1998). The single cells were aspirated with a micropipette (diameter of the pipette tip ~1.5 µm) and at temperatures lower than body temperature all cells blocked the pipette. Above body temperature all aspirated RBC passed the narrow micropipette tip easily without any apparent resistance. The passage temperature was 36.3 ± 0.3 °C being remarkably close to human body temperature (Artmann et al. 1998). It was found that the passage behavior is caused by a reduction of the viscosity of the concentrated Hb solution in the RBC (Artmann et al. 1998). The loss of viscosity and the passage transition in the micropipette experiments were found to be connected to perturbations and partial unfolding of the structure of Hb at body temperature (Artmann et al. 2004). Further studies revealed that the structural perturbation of Hb at body temperature leads to Hb aggregation above ~37 °C (Digel et al. 2006), and concomitantly RBC release cytosolic cell water to the outside blood plasma as observed in colloid osmotic pressure measurements (Artmann et al. 2009). A direct correlation between the structural perturbation temperature of Hb and the body temperature of a large variety of different species was reported, which further supported the biological relevance of the effect (Digel et al. 2006; Zerlin et al. 2007). It was speculated that the partial loss of Hb structure causes an increase in surface hydrophobicity, which might result in stronger protein-protein interactions and thus lead to protein aggregation above body temperature (Digel et al. 2006; Stadler et al. 2008a; Stadler et al. 2009).

Krueger and co-authors studied the interactions of Hb in RBC and concentrated solution and demonstrated that a hard sphere potential plus screened electrostatics can approximately describe the protein-protein interaction potential (Krueger et al. 1990; Krueger and Nossal 1988). The same results was obtained later also for concentrated myoglobin solutions (Longeville et al. 2003). It might be of interest to note that studies on concentrated solutions of crystallins and of lysozyme demonstrated that a delicate balance between hard sphere and weak attractive interactions are crucial for the stability of these concentrated protein solutions. (Cardinaux et al. 2007; Dorsaz et al. 2009; Stradner et al. 2007). In further experiments Doster and Longeville examined the diffusion of Hb in RBC using neutron spin-echo spectroscopy (Doster and Longeville 2007). The authors had the idea to interpret the diffusion of Hb in RBC using the theory of colloidal diffusion at high concentration. The neutron spin-echo technique is sensitive to molecular motions occurring in the *ns* and *nm* time- and length-scale. Doster and Longeville compared the measured short-time and long-time self-diffusion coefficient of Hb to theoretical calculations of non-charged hard sphere suspensions with direct and hydrodynamic interactions (Doster and Longeville 2007). It was necessary to include the hydration shell as a hydrodynamic coat to release the discrepancy with colloidal theory. Furthermore, it was deduced that hydrodynamic and not direct interactions dominate Hb diffusion at high concentration.

Without hydration water, proteins would neither fold correctly (Chaplin 2006; Cheung et al. 2002; Dobson et al. 1998) nor acquire the conformational flexibility, which is considered relevant for biological activity (Rupley and Careri 1991). Motions in proteins occur over a very large range of time-scales from fast reorientations of amino acid side chains in the *ps* range, to slower motions of the protein backbone in the *ns* time-scale and to very slow processes of protein subunits and folding processes in the *µs* and *ms* range (McCammon and Harvey 1987). Fast fluctuations in the *ps* and *ns* time-scale are considered to act as lubricant and to enable much slower physiologically important motions (Brooks et al. 1988). Hydration dependent internal protein dynamics has been studied with incoherent neutron scattering in several model systems mainly as hydrated powders, including myoglobin (Doster 2008; Doster et al. 1989), lysozyme (Cornicchi et al. 2005; Marconi et al. 2008; Paciaroni et al. 2005), and α-amylase (Fitter 1999, 2003; Fitter and Heberle 2000). In incoherent neutron scattering experiments, the single particle motions of hydrogen (H) atoms are detected. H atoms are indicators of average protein dynamics as they constitute ~50% of the atoms and are uniformly distributed in the macromolecules (Gabel et al. 2002). Hydration water not only enables protein dynamics but participates actively in protein function. Around 60 additional water molecules are bound in the hydration layer of the oxygenated form of Hb as compared to the deoxygenated state of Hb (Colombo et al. 1992). The additional water molecules were found to be thermodynamically important for regulation of Hb activity. The study of Colombo and coworkers was done in aqueous solution at a Hb concentration of 64 mg/ml. Further studies revealed that binding of the extra water molecules is the rate-limiting step of Hb activity (Salvay et al. 2003). Therefore, it is an important question if protein dynamics is adapted to the specific hydration level in cells.

In this manuscript, we present a study of Hb dynamics in RBC in the *ps* to *ns* time- and *Å* length-scale using high-resolution quasielastic neutron scattering (QENS). The aim of the study is to demonstrate how QENS allows the measurement and separation of internal protein dynamics and global macromolecular diffusion in whole cells. The QENS technique provides complementary information to fluorescent correlation spectroscopy (Schwille et al. 1999; Wawrezinieck et al. 2005) or neutron spin-echo spectroscopy (Doster and Longeville 2007; Lal et al. 2010; Le Coeur and Longeville 2008) which are sensitive to different time-space windows of protein fluctuations.

# Material and Methods

## *Sample preparation*

Samples of human venous blood from healthy adults were drawn with tubes containing heparin to prevent blood coagulation. RBC samples were prepared as described in Stadler et al. (Stadler et al. 2008a). During the sample preparation, the RBC were gassed with CO to increase the stability of Hb and the glycocalyx matrix was removed. The cells were washed several times with $D_2O$ HEPES buffer (137 mM NaCl, 4 mM KCl, 1.8 mM $CaCl_2$, 0.8 mM $Na_2HPO_4$, 0.2 mM $NaH_2PO_4$, 0.7 mM $MgSO_4$, 8.4 mM HEPES, and 4 mM NaOH) at pD=7.4 and 290 mOsm to reduce the neutron scattering contribution of the buffer. The washing steps were repeated until the level of $H_2O$ was estimated to be below 0.1 vol%. The shape of the cells was checked with optical microscopy after the washing steps. The cell pellet was sealed in a flat aluminum sample holder of 0.2 mm thickness for the neutron scattering experiment. It was checked by weighing that there occurred no loss of sample material during the experiment.

## *Dynamic light scattering experiments*

Samples for the dynamic light scattering experiments were prepared from a blood drop taken from the finger tip. The RBC were washed with $H_2O$ HEPES buffer and lysed with distilled water. The sample for dynamic light scattering experiments was not gassed with CO. Before the dynamic light scattering experiments, the dilute Hb solution in $H_2O$ buffer (0.1M KCl, 61.3 mM $K_2HPO4$, 5.33 mM $KH_2PO4$, pH 7.4, 290-300 mOsm) was centrifuged at 20000 relative centrifugal force and filtered using 0.25 µm nitrocellulose filters. UV/VIS absorption spectroscopy was used to determine the concentration of the Hb solution. The Hb was found to be in the oxy-state as evidenced by the characteristic bands in the absorption spectrum, and the protein concentration was 0.4 mg/ml. The protein concentration was determined using extinction coefficients of 13.8 mM$^{-1}$*cm$^{-1}$ at 541 nm and 128 mM$^{-1}$*cm$^{-1}$ at 405 nm for oxy-Hb, the molar concentration is per heme group (Antonini and Brunori 1970). Dynamic light scattering of dilute human Hb solution was measured on a Wyatt DAWN-EOS instrument (Wyatt Technology, Santa Barbara, CA) and corrected for temperature dependent D$_2$O viscosity using literature values (Cho et al. 1999). The diffusion coefficients were calculated using the ASTRA 5 software package from the manufacturer. Around 5 ml of sample was measured per experiment.

## *Neutron scattering experiments*

Neutron scattering was measured on the high-resolution neutron backscattering spectrometers IN10 and IN16 at the ILL (http://www.ill.eu/instruments-support/instruments-groups/yellowbook/) and on IRIS at the ISIS spallation source (http://www.isis.stfc.ac.uk/instruments/iris/). To minimize multiple scattering, RBC samples with high transmissions were used (0.95 on IN16 and IRIS, 0.9 on IN10). The instruments IRIS, IN10 and IN16 are characterized by energy resolutions $\Delta E$ of 17, 1 and 0.9 µeV (FWHM), respectively, which correspond to slowest observable motions in the order of $\Delta t = \hbar / \Delta E$ ~40 ps and ~1 ns, respectively. Neutron scattering was measured in the range of $0.49 \leq q \leq 1.6$ Å$^{-1}$ on IN16, $0.5 \leq q \leq 1.45$ Å$^{-1}$ on IN10 and $0.48 \leq q \leq 1.6$ Å$^{-1}$ on IRIS, where $q$ is the modulus of the scattering vector. The instrumental energy resolution was determined with a vanadium measurement. The scattering contribution of the empty aluminum sample holder was subtracted from the measured data. Neutron detectors were grouped on IN16 and IRIS to obtain better statistics. Incoherent scattering of D$_2$O solvent contributes partially to the measured intensities: Free and interfacial water dynamics are out of the *Å-ns* space and time window of IN10 and IN16 and contribute only as a flat background to the measured spectra (Tehei et al. 2007). Experimental data is dominated by Hb motions on the IRIS spectrometer, and the incoherent contribution of D$_2$O on IRIS is estimated to be smaller than 4% at $q$<1.3 Å$^{-1}$ (Stadler et al. 2008a). Gaspar and coworkers evaluated the coherent and incoherent scattering contributions of concentrated protein solutions in D$_2$O solvent (Gaspar et al. 2010). In a completely dry myoglobin powder the authors found an incoherent scattering fraction of ~90% and a coherent scattering fraction of ~10% between 0.5 and 1.5 Å$^{-1}$. For a concentrated myoglobin solution of 360 mg/ml, the authors reported an incoherent scattering fraction of around 80% and a coherent scattering fraction of around 20% between 0.5 and 1.5 Å$^{-1}$. The coherent scattering fraction of D$_2$O in the 360 mg/ml solution therefore has to be ~10% in that scattering vector range. In RBC the protein concentration is 330 mg/ml and the values should be comparable.

## QENS data analysis

The scattering function of internal protein dynamics $S_I(q,\omega)$ can be written in simplified form as an elastic term and a single Lorenzian that represents internal protein diffusive motions (Gabel et al. 2002)

$$S_I(q,\omega) = A(q) \cdot \delta(\omega) + (1 - A(q)) \cdot \frac{1}{\pi} \cdot \frac{\Gamma_I(q)}{\omega^2 + \Gamma_I(q)^2} \,. \tag{1}$$

The prefactor *A(q)* is called *Elastic Incoherent Structure Factor* (EISF), *q* is the modulus of the scattering vector, and *Γ$_I$(Q)* are the Half-Widths at Half-Maximum (HWHM) of the Lorentzian. The *q*-dependence of the EISF contains information about the geometry of localized motions, and the scattering vector dependence of *Γ$_I$(q)* informs about the diffusion coefficients and residences times of diffusive motions.

Global macromolecular diffusion consists of translational and rotational diffusion of the protein. The scattering function of global protein diffusion $S_G(q,\omega)$ is the convolution of the scattering functions of translational and rotational diffusion assuming that rotational and translational diffusion are uncorrelated. It was shown theoretically by Perez and co-workers that rotational diffusion of a protein leads to an additional broadening of the measured HWHM (Perez et al. 1999). The scattering function $S_G(q,\omega)$ could be approximated by a single Lorentzian with the half-widths *Γ$_G$(q)*,

$$S_G(q,\omega) = \frac{1}{\pi} \cdot \frac{\Gamma_G(q)}{\omega^2 + \Gamma_G(q)^2} \;. \tag{2}$$

The line-widths and the apparent diffusion coefficient $D_{app}$ of the protein are related by $\Gamma_G(q) = D_{app} \cdot q^2$ (Perez et al. 1999). The apparent diffusion coefficient $D_{app}$ was compared to $D_0$, which is the translational diffusion coefficient of the protein at infinite dilution. The identical value of $D_{app}/D_0 = 1.27$ was obtained for myoglobin and hemoglobin (Perez et al. 1999; Stadler et al. 2008a). The calculation of the contributions of rotational and translational diffusion to the measured spectra is described in the appendix.

Furthermore, it is assumed that internal protein dynamics and global protein diffusion are uncorrelated in concentrated protein solutions. The scattering function $S(q,\omega)$ then is the convolution between $S_I(q,\omega)$ and $S_G(q,\omega)$, $S(q,\omega) = S_I(q,\omega) \otimes S_G(q,\omega)$ (Bee 1988). The scattering function reads as

$$S(q,\omega) = \exp(-<x^2>q^2) \cdot \left\{ \frac{A(q)}{\pi} \cdot \frac{\Gamma_G(q)}{\Gamma_G(q)^2 + \omega^2} + \frac{1-A(q)}{\pi} \cdot \frac{\Gamma_G(q) + \Gamma_I(q)}{[\Gamma_G(q) + \Gamma_I(q)]^2 + \omega^2} \right\} \tag{3}$$

where the exponential represents a Debye-Waller factor for fast molecular vibrations. $S(q,\omega)$ plus linear background was convoluted with the instrumental resolution function and fitted to the measured QENS spectra in the energy range of $-14 \leq E \leq +14\,\mu eV$ for IN16, $-12.4 \leq E \leq +12.4\,\mu eV$ for IN10 and $-0.5 \leq E \leq +0.5\,meV$ for IRIS using the DAVE software package (Azuah et al. 2009).

Gaspar and coworkers demonstrated that the half-widths of internal motions $\Gamma_I(q)$ are a weaker parameter compared to the EISF (Gaspar et al. 2008). The authors could fit measured QENS spectra equally well with constant or freely varying line-widths as a function of the scattering vector. As a test we fixed the line-widths of internal motions to the $q$-independent average value of $\Gamma_I(q)=0.2\,meV$. The obtained line-widths $\Gamma_G(q)$ of global Hb diffusion were then found to increase linearly with $q^2$ as expected but did not intercept zero at $q^2 \rightarrow 0$. A non zero intercept at $q^2 \rightarrow 0$ of global protein diffusion is difficult to interpret with global Hb diffusion. On the other hand, when the line-widths $\Gamma_I(q)$ were allowed to vary freely, we obtained line widths $\Gamma_G(q)$ that pass through zero as expected for global protein diffusion.

# Results and Discussion

In the following we present and discuss the results of our experiments. Typical QENS data measured on the neutron spectrometers IN16, IN10, and on IRIS are shown in Figure 1. The measured spectra were well described with a narrow and a broad Lorentzian for global macromolecular diffusion and internal Hb dynamics, respectively. First, we discuss the results about global Hb diffusion. Our interpretation follows the ideas of Doster and Longeville (Doster and Longeville 2007).

## *Global macromolecular diffusion*

The measured half-widths at half-maximum (HWHM) for global Hb diffusion are presented in Figure 2. Apparent diffusion coefficients $D_{app}$ were determined according to $\Gamma_G(q) = D_{app} \cdot q^2$ in the range of $0.24 \leq q^2 \leq 2.56$ Å$^{-2}$ for IN10 and IN16 data, and in the range of $0.72 \leq q^2 \leq 2.57$ Å$^{-2}$ for IRIS data. The line-widths $\Gamma_G(q)$ of global Hb diffusion increase linearly with $q^2$ up to around 2.6 Å$^{-2}$. This behavior is a clear sign for continuous global diffusion of Hb. The $D_{app}$ contain both a component of translational and rotational diffusion of Hb. It was shown previously (Perez et al. 1999; Stadler et al. 2008a) that rotational diffusion of Hb leads to an additional broadening of the spectra by the factor 1.27. Therefore, the apparent diffusion coefficients $D_{app}$ were divided by 1.27 to obtain the global translational diffusion coefficient $D$ of Hb. The essential steps in the calculation of the contributions of rotational and translational diffusion to the experimental spectra are outlined in the appendix. All obtained values of $D$ are compared in Figure 3. The line-widths obtained from the measurements with IN16 and IN10 intercept zero (Figure 2 a), which indicates that on time-scales of ~1 ns global Hb diffusion does not sense confinement of neighboring proteins. The $\Gamma_G(q)$ measured on IRIS appear to converge towards a plateau at small $q^2$ and low temperature (Figure 2 b). The feature indicates a cage effect of the neighboring molecules on Hb diffusion in the *ps* time-scale and was observed before (Stadler et al. 2008a). Multiple scattering might lead to a deviation from linear behavior at small $q^2$. However, as the transmission of the sample was 0.95 multiple scattering should be completely negligible. An alternative explanation could be that small uncertainties of the resolution function might result in a plateau at small $q^2$, as the HWHM are only ~10% of the energy resolution of IRIS. An observation time dependent diffusion coefficient is obtained. Unruh and coworkers observed a similar phenomenon (Unruh et al. 2008). The authors studied the motions in liquid medium-chain n-alkanes using QENS with observation times from 1.1 ps to 900 ps and molecular dynamics simulations. The study revealed a time dependent diffusion coefficient, and there was no need to use the obtained half-widths at low-resolution for the analysis of the high-resolution data. To check the validity of our interpretations we have also performed a complementary analysis in time-space (see Supplementary Material). The obtained diffusion coefficients in time- and in energy-space are identical within the error bars. In the case that the diffusion coefficient measured with IRIS would be visible with IN16/IN10, we would obtain a mixture in time-space of the IRIS and the IN16/IN10 energy-space results. This is not the case and our check therefore demonstrates the validity of our analysis.

Studies on average macromolecular dynamics in *E. coli* cells (Jasnin et al. 2008) and in concentrated myoglobin solutions (Busch et al. 2007) using high-resolution neutron backscattering spectroscopy reported that the measured line-widths of global macromolecular diffusion deviate from linear behavior and tend towards saturation at large $q^2$. Jump-diffusion of the macromolecules was discussed as a possible explanation (Busch et al. 2007; Jasnin et al. 2008), as this mechanism would result in a saturation of the line-widths at large $q^2$. On the other hand, a distribution of diffusion coefficients could also be responsible for the deviation of the line-widths from linear behavior (Busch et al. 2007). Importantly, any kind of non-localized diffusion leads to line-widths that tend towards zero with $\Gamma_G(q) = D_{app} \cdot q^2$ at small $q^2$-values. Jasnin and co-workers studied average macromolecular dynamics in *E. coli* using the IRIS spectrometer (Jasnin et al. 2008). Global macromolecular diffusion in *E. coli* was too slow and could not be resolved with IRIS. Prokaryotic cells, such as *E. coli*, are very complex objects which contain a vast amount of large macromolecular assemblies, such as ribosomes with a molecular mass of 2.5 MDa. Average macromolecular dynamics in *E. coli* are therefore difficult to attribute to a certain component. Hb is the main macromolecular component of RBC with a rather small molecular mass of 65 kDa. It is reasonable to assume that global diffusion of Hb is significantly faster than that of large macromolecular complexes in *E. coli*, which would explain why Hb global diffusion in RBC is visible on IRIS.

Tokuyama and Oppenheim evaluated the short-time $D_S^S$ and long-time $D_S^L$ self-diffusion coefficients of concentrated non-charged hard-sphere suspensions with hydrodynamic and direct interactions as a function of the volume fraction $\phi$ and of the diffusion coefficient at infinite dilution $D_0$ (Tokuyama and Oppenheim 1994). Short-time self-diffusion corresponds to particles that move in a static configuration of the neighboring particles at times $t<\tau_D$, with the structural relaxation time $\tau_D$. The long-time limit of self-diffusion is reached at $t>\tau_D$. The values of the short- and long-time self-diffusion coefficients are equal only in dilute solution. At higher concentrations short-time self-diffusion is always faster than long-time self-diffusion. We measured the diffusion coefficient $D_0$ of Hb at infinite dilution with dynamic light scattering. The theoretical values of $D_S^S$ and $D_S^L$ of Hb at a volume fraction of $\phi=0.25$ ($D_S^S = 0.56 \cdot D_0$, $D_S^L = 0.28 \cdot D_0$) are given in Figure 3 (Tokuyama and Oppenheim 1994). It is obvious that the measured diffusion coefficients are too small and do not agree with the theoretical values. Full hydration of myoglobin corresponds to a value of $h$~0.39g H$_2$O/ g Mb (Rupley and Careri 1991). It is believed that the critical hydration to allow the onset of anharmonic motions in myoglobin (Mb) is around $h_{Mb}$=0.35g H$_2$O/g Mb (Doster et al. 1989). Hb has a larger radius of gyration than Mb (Hb: $R_G$~24Å, Mb: $R_G$~16Å) and a smaller surface to volume ratio *S/V* (Longeville et al. 2003; Schelten et al. 1972). Approximating Hb and Mb as spherical particles, the critical hydration of Hb should be around $h_{Hb}$=(S/V)$_{Hb}$ / (S/V)$_{Mb}$*$h_{Mb}$=16Å/24Å*0.35g H$_2$O/ g Mb=0.23g H$_2$O/ g Hb. It is known that the density of protein hydration water is ~10% larger than bulk solvent (Svergun et al. 1998). The partial specific volume ν of Hb plus hydration water is then $v = (0.75 + 0.23/1.1)\, ml/g = 0.98\, ml/g$, which corresponds to an effective volume fraction of Hb plus hydration water of $\phi = c \cdot v = 0.32$ with the concentration c=0.33 g/ml of Hb in RBC (DeMoll et al. 2007; Doster and Longeville 2007). The measured self-diffusion coefficients of Hb with IN16, IN10 and IRIS agree with high accuracy with the theoretical values of Hb plus hydration shell ($D_S^S = 0.45 \cdot D_0, D_S^L = 0.18 \cdot D_0$) (Tokuyama and Oppenheim 1994). In a study on the short-time limit of Hb diffusion in RBC we estimated that the structural relaxation time $\tau_D$ is in the order of several hundred ps (Stadler et al.

2008a). Therefore, IRIS is sensitive to motions which are faster than $\tau_D$ and short-time self-diffusion is detected. The high-resolution instruments IN16 and IN10, and neutron spin echo spectroscopy (Doster and Longeville 2007) measure motions which are longer than $\tau_D$, and the long-time limit of Hb self-diffusion is observed. We showed previously that although ~90% of cell water in RBC has properties similar to bulk water, a small fraction of ~10% cellular water exhibits strongly reduced dynamics and was attributed to water molecules which are bound to the surface of Hb (Stadler et al. 2008b). The ratio of water per Hb in RBC is $h$~2.3g $H_2O$/ g Hb and a ~10% fraction corresponds to a bound water fraction of ~0.23g $H_2O$/ g Hb which is identical to the value reported in this article. Furthermore, Doster and Longeville measured the diffusion of Hb in whole red blood cells using spin echo spectroscopy in the *ns* and *nm* time- and length-scale (Doster and Longeville 2007). The authors demonstrated that the presence of the hydration shell leads to a reduction of the diffusion coefficient of Hb. Garcia de la Torre calculated hydrodynamic properties of proteins from atomic structures (Garcia de la Torre 2001). It was demonstrated by comparing calculated and experimental values that a hydration shell of $h$~0.27g $H_2O$/ g Hb is bound to the surface of Hb. Our result is reasonably close to that value.

## *Internal hemoglobin dynamics*

We now turn our attention to the results concerning internal protein dynamics. Detailed information about protein internal motions can be extracted from the scattering vector dependence of the quasielastic broadening and the *Elastic Incoherent Structure Factor* (EISF). The HWHM of internal protein dynamics measured on IN16 and on IRIS are given in Figure 4 (a) and (b). The $\Gamma_I(q)$ measured with IRIS show typical behavior of localized jump-diffusion. The half-widths tend towards a constant value at small $q^2$, which indicates confining effects of local boundaries. Diffusive jumps with a finite jump-length lead to a plateau in the line-widths at large scattering vectors. In the $q^2$-range of 0.72 and 2.57 Å$^{-2}$ the $\Gamma_I(q,\omega)$ measured on IRIS could be well described with a jump-diffusion model given by $\Gamma_I(q,\omega) = \dfrac{D_I q^2}{1 + D_I q^2 \tau}$ (Bee 1988). The parameters of the jump-diffusion model are the residence time before a jump $\tau$ and the jump-diffusion coefficient $D_I$ of protein internal motions. In Figure 5, the jump-diffusion coefficients and the residence times of internal Hb dynamics in RBC are compared to jump-diffusion coefficients and residence times of internal Hb dynamics as hydrated powder ($h$=0.4 g $D_2O$/ g Hb) and as concentrated solution ($h$=1.1 g $D_2O$/ g Hb). The corresponding hydration level in RBC is $h$~2.5 g $D_2O$/ g Hb. The results of the experiments with the hydrated Hb powder and concentrated Hb solution have been published before and are given here for comparison (Stadler et al. 2009). The Hb powder and solution were measured on neutron time-of-flight spectrometers with energy resolutions of 50 and 100 µeV, respectively. All data were analyzed in the same way. The results demonstrate that an increase in the hydration level from one hydration shell in the Hb powder to around 3 hydration layers in the concentrated Hb solution increases the jump-diffusion coefficients and strongly reduces the residence times of internal protein dynamics in the *ps* time-scale. A further increase in the hydration level to around 6 hydration layers per Hb in whole RBC does neither enhance the jump-diffusion coefficients nor reduce significantly the residence times as compared to the concentrated Hb sample. The rate of internal jump-diffusion in the *ps* time-scale appears to be already fully developed in the concentrated Hb solution. The observed motions in the *ps* range could correspond to diffusive jumps of amino acid side chains and attached methyl groups (Fitter et al. 1996).

The half-widths of internal protein dynamics from the experiment on IN16 are independent of the scattering vector within the error bars, as shown in Figure 4(a), and have average values of 5.8 ± 1.4 µeV at 11.9 °C and 6.2 ± 1.0 µeV at 26.9 °C. The line-widths determined with IN10 are 5.5 µeV at 19.1 °C and 4.2 µeV at 36.5 °C. The $\Gamma_I(q)$ on each individual spectrum of IN10 had large errors. The average value was used for all spectra at one temperature and held constant during fitting of IN10 data; the obtained values are rather imprecise and are given only for completeness. The line-widths obtained on IN16 and IN10 are in agreement with other studies which investigated protein dynamics in the *ns* time-scale using high-resolution quasielastic neutron scattering. Fitter and co-workers studied hydrated bacteriorhodopsin and obtained half-widths of 5.5 µeV (Fitter et al. 1997), Orecchini and co-workers investigated hydrated β-lactoglobulin powder and found half-widths of 16 µeV (Orecchini et al. 2002), Busch et al. found line-widths of 10 µeV of myoglobin in concentrated solution (Busch et al. 2007), and Jasnin et al. measured average dynamics in whole *E. coli* and obtained line-widths of ~7 µeV (Jasnin et al. 2008). If we exclude the lactoglobulin case, the values of the measured line-widths are rather similar although the hydration levels in the investigated systems are different. We recall that correlation times τ and line-widths Γ are inversely related by $\tau_{cor} = 1/\Gamma$. This seems to indicate that correlation times of motions in globular and membrane proteins in the *ns* time-scale are rather similar in hydrated protein powders, solutions and in whole cells. As the observed line-widths on IN16 and IN10 are independent of the scattering vector, a different class of motions is observed using the high-resolution instruments. Rotational motions lead to line-widths which are independent of the scattering vector (Bee 1988), and the observed dynamics might be attributed to slow rotations of side chains or relaxations of the protein backbone.

Information about the geometry of motions can be extracted from the measured EISF. Only four and six data points are available on IN10 and IN16, respectively. This is too few and does not allow an accurate analysis. Therefore, we limit our discussion to the results of the experiment using the IRIS spectrometer. The EISF obtained with IRIS at 26.9 °C is shown in Figure 6. The EISF was interpreted with the model of Volino and Dianoux for diffusion in a sphere (Volino and Dianoux 1980). The diffusion in a sphere model can be written as $A_0(q) = p + (1-p) \cdot \left[ \frac{3 j_1(qa)}{qa} \right]^2$, where $j_1(qa)$ is the first-order spherical Bessel function of the first kind, *a* is the sphere radius, and $A_0(q)$ is the EISF. The hydrogen atoms which appear immobile and mobile within the instrumental energy resolution are represented by the fractions *p* and *(1-p)*, respectively. The obtained sphere radius *a* increases from 2.8 ± 0.1 Å at 16.9 °C to 3.3 ± 0.1 Å at 36.9 °C. The immobile fraction *p* has got the average value of 0.67. These values reasonable agree with results on macromolecular dynamics in *E. coli* which found *a*=3.1 Å and *p*=0.61 at 6.9 °C; *a*=3.4 Å and *p*=0.56 at 26.9 °C (Jasnin et al. 2008).

To take into account of the heterogeneity of internal protein dynamics Perez and co-workers extended the diffusion in a sphere model and introduced a Gaussian distribution of sphere radii *f(a)* instead of a single sphere (Perez et al. 1999). The Gaussian distribution is defined as $f(a) = \frac{2}{\sigma\sqrt{2\pi}} \exp(-a^2/2\sigma^2)$, with the standard deviation σ as free parameter. The mean value of the sphere radius is given by $\hat{a} = \sigma\sqrt{\frac{2}{\pi}}$. A neutron scattering study using specific isotope labeling in order to investigate the dynamics of specific amino acids in bacteriorhodopsin demonstrated the heterogeneity of internal protein dynamics (Wood et al. 2008). The obtained average sphere radius $\hat{a}$ increases from $\hat{a}$=2.1 ± 0.1 Å at 16.9 °C to $\hat{a}$=3.0 ± 0.2 Å at 36.9 °C.

The immobile fraction *p* has an average value of 0.50 and increases only slightly with temperature from *p*=0.47 ± 0.02 at 16.9°C to *p*=0.55 ± 0.01 at 36.9°C. Using the same model we have quantified the average amplitudes of motion in concentrated Hb solution (Stadler et al. 2009). The average sphere radius was found to increase from *â*=2.3 Å at 6.9 °C to *â*=2.6 Å at 36.9 °C, while the immobile fraction was constant with temperature *p*=0.38 (Stadler et al. 2009). Within the error bars the obtained average sphere radii of Hb in RBC and of Hb in concentrated solution are similar when we exclude the value at 36.9 °C, which is larger in Hb in RBC than in the Hb solution. Although the energy resolutions of the instruments used for both experiments are different (17μeV and 100μeV, respectively) the observed motions look similar. This might either be due to the fact that the same motions are seen using the IRIS and the time-of-flight spectrometers, or that different classes of motions in the order of 40ps and several ps are similar. The second possibility would also imply that the corresponding hierarchical structures in the energy landscape are similar.

The model for diffusion in a sphere approximately describes the measured EISF. Better fits can be obtained when a Gaussian distribution of sphere radii is used. It should be noted that both the diffusion in a sphere model and the Gaussian distribution can only be simple and rough representations for the heterogeneity of internal protein dynamics. In any case, models are never wrong, they are just more or less appropriate.

# Conclusion

In summary, we measured the global self-diffusion and internal dynamics of Hb in RBC, *in vivo*, using high-resolution quasielastic neutron backscattering spectroscopy. It is demonstrated that global protein diffusion and internal dynamics can be separated and interpreted quantitatively. The cross-over from the short to the long time limit of Hb self-diffusion could be observed. It is demonstrated that the diffusion of Hb at high concentration in RBC can be described with concepts of colloid physics. Experimental data is in quantitative agreement with hydrodynamic theory of non-charged hard-sphere suspensions when it is assumed that the hydration shell moves with the protein. It is shown that interfacial protein hydration water has a strong influence on global protein diffusion under physiological conditions in cells. The same result was obtained by Doster and Longeville using spin-echo spectroscopy (Doster and Longeville 2007). Experiments with whole RBC using micropipette aspiration and colloidal osmotic pressure measurements (Artmann et al. 2009; Artmann et al. 1998) indicated that the cellular environment might have similarities to a colloidal gel. It was suggested that the trigger for the formation of the gel could be Hb-Hb interactions, which are influenced by the molecular properties of Hb (Digel et al. 2006; Zerlin et al. 2007). Recently, we studied Hb-Hb interactions in concentrated solution using small angle neutron scattering and could show that Hb molecules associate into a large-scale superstructure at high concentration (Stadler et al. 2010). In this article, we observe a slowing down of the atomistic diffusion of Hb, which might indeed lead to gel-like properties on a macroscopic scale. It is demonstrated how incoherent neutron scattering can contribute to the understanding of cellular phenomena on a macroscopic scale.

Internal Hb dynamics was also measured and could be separated from global Hb diffusion. The internal motions of Hb were compared to results obtained with hydrated powder and solution samples. Different types of motions were brought into focus by using neutron spectrometers with specific energy resolutions. Hydration water was found to have a strong influence on motions in the *ps* time-scale. Jump-diffusion coefficients of internal Hb fluctuations are significantly enhanced and residence times of the internal diffusive jumps are reduced in RBC as compared to fully hydrated Hb powder. Slower internal dynamics of Hb in RBC in the *ns* time-range were found to be rather similar to results obtained with fully hydrated protein powders, solutions and *E. coli* cells. Still missing is a combined analysis of the data measured with different spectrometers, which should be done in a future publication. Future work might also be dedicated to investigate protein dynamics in whole cells under different environmental conditions.

# Acknowledgements

The author (A.M.S.) thanks Georg Büldt for continuous support. We also thank Giuseppe Zaccai for valuable discussion and critical reading of the manuscript.

# Appendix

# Global Hb diffusion: Contribution of rotational and translational diffusion

Global protein diffusion consists of translational and rotational protein diffusion around the center of mass. Free translational diffusion is described by the scattering function

$$S_{trans}(q,\omega) = \frac{1}{\pi} \cdot \frac{\Gamma_{trans}(q)}{\omega^2 + \Gamma_{trans}(q)^2} ,\qquad (A\ 1)$$

with the diffusion coefficient $\Gamma_{trans}(q) = D_0 \cdot q^2$ (Bee 1988). It was shown theoretically by Perez and co-workers that rotational diffusion of a protein leads to an additional broadening of the measured HWHM (Perez et al. 1999). Rotational and translational diffusion of the protein are assumed to be uncorrelated. In that case the scattering function of global protein diffusion $S_G(q,\omega)$ is the convolution of the scattering functions of translational and rotational diffusion

$$S_G(q,\omega) = S_{trans}(q,\omega) \otimes S_{rot}(q,\omega) = \frac{1}{\pi} \cdot \sum_{l=0}^{\infty} B_l(q) \cdot \frac{\Gamma_{trans}(q) + \Gamma_l(q)}{\omega^2 + [\Gamma_{trans}(q) + \Gamma_l(q)]^2} ,\qquad (A\ 2)$$

with $\Gamma_l = l(l+1) \cdot D_{rot}$ and the rotational diffusion coefficient $D_{rot}$ (Perez et al. 1999). The integrals in the terms $B_0(q)$ and $B_l(q)$ are extensions of the Sears model (Sears 1966) for rotation on the surface of a sphere. They describe the distribution of hydrogen atoms within the protein

$$B_0(q) = \int_{r=0}^{R} 4\pi r^2 \cdot j_0^{\,2}(qr)\,dr, \quad B_{l\geq 1}(q) = \int_{r=0}^{R} 4\pi r^2 \cdot (2l+1) \cdot j_l^{\,2}(qr)\,dr. \tag{A 3}$$

The terms $j_l$ are the $l^{th}$-order spherical Bessel function of the first kind and $R$ is the radius of the Hb. The terms $B_l(q)$ were integrated numerically and the obtained scattering function $S_G(q,\omega)$ could be perfectly approximated by a single Lorentzian with the apparent diffusion coefficient $D_{app}$ and the HWHM $\Gamma_G(q) = D_{app} \cdot q^2$. The apparent diffusion coefficient $D_{app}$ was compared to $D_0$, which gave the relation of $D_{app}/D_0 = 1.27$ (Perez et al. 1999; Stadler et al. 2008a).

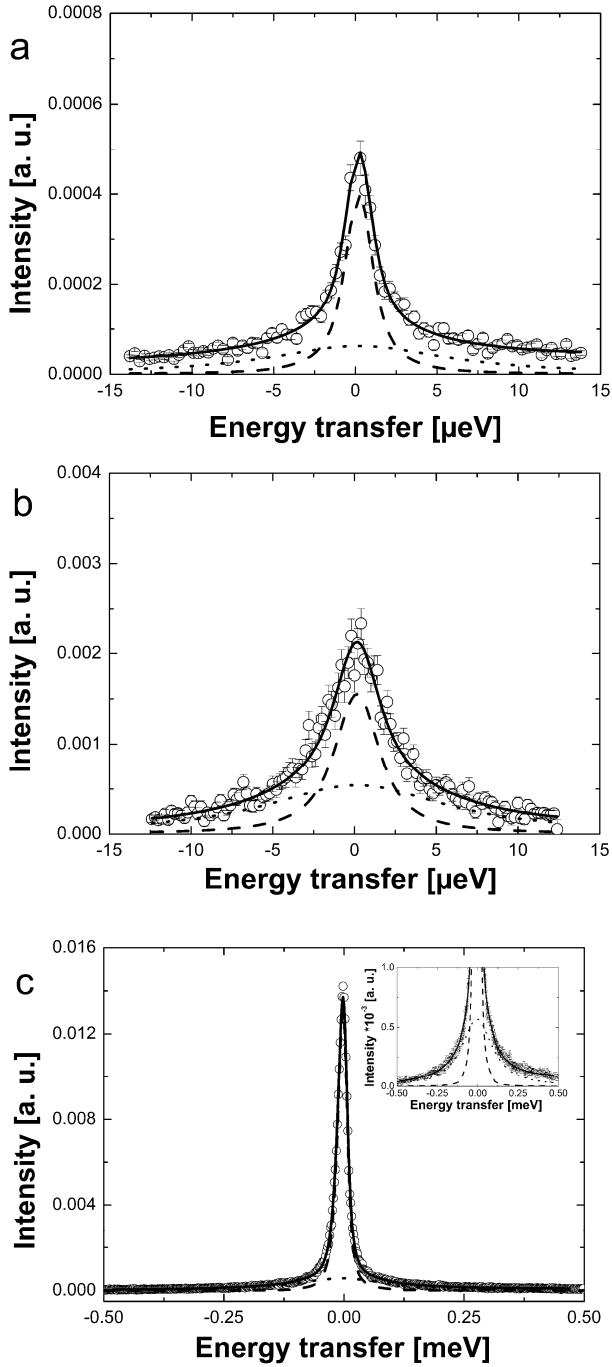

**Figure 1:** Experimental QENS data of Hb in RBC measured on (a) IN16 at 11.9°C and $q$=1.3 Å$^{-1}$, (b) IN10 at 19.1°C and $q$=1.45 Å$^{-1}$, (c) IRIS at 16.9°C and $q$=1.37 Å$^{-1}$. The solid black line is the total fit, the dashed and the dotted lines represent the narrow and broad Lorentzians used for data analysis. The inset in (c) on the right side shows a magnification of the spectrum measured on IRIS to illustrate the quality of the fit. The instruments IRIS, IN10 and IN16 are characterized by energy resolutions $\Delta E$ of 17, 1 and 0.9 µeV (FWHM), respectively.

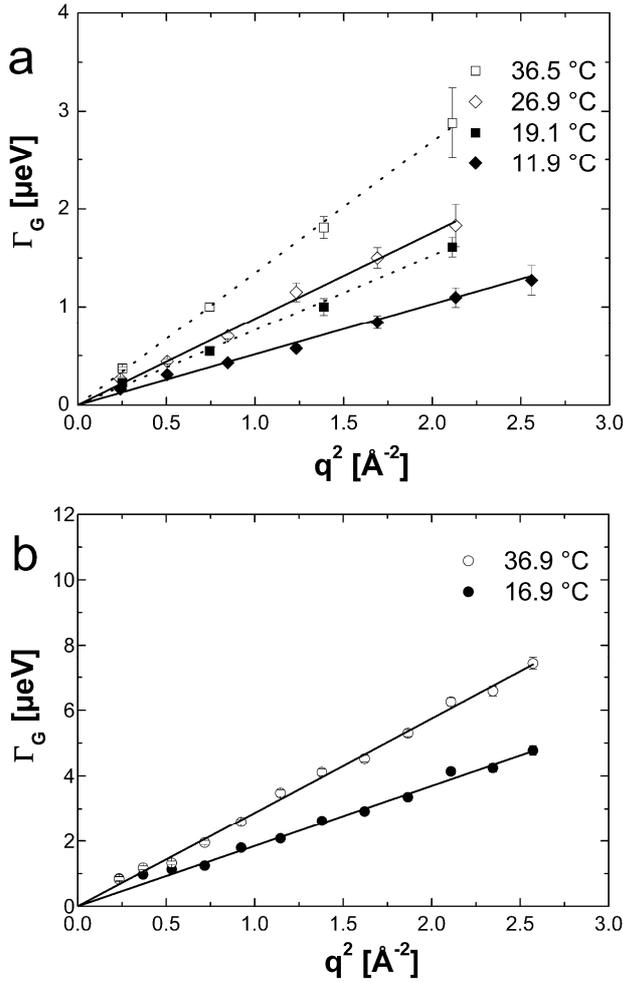

**Figure 2:** Half-widths at half-maximum of the Lorentzians of global Hb diffusion in RBC measured with QENS on (a) IN16 (diamonds), IN10 (squares) and (b) IRIS. The straight lines are linear fits to the data. In (a) the solid lines are fits to IN16 and the dotted lines are fits to IN10 data. The linear increase of the line-widths with $q^2$ is a clear sign for continuous global Hb diffusion. The diffusion coefficients of Hb were determined from the slope of the linear fits. IN10 and IN16 are sensitive to motions in the time-scale of ~*ns*, whereas IRIS detects motions in the time-scale of ~40 *ps*.

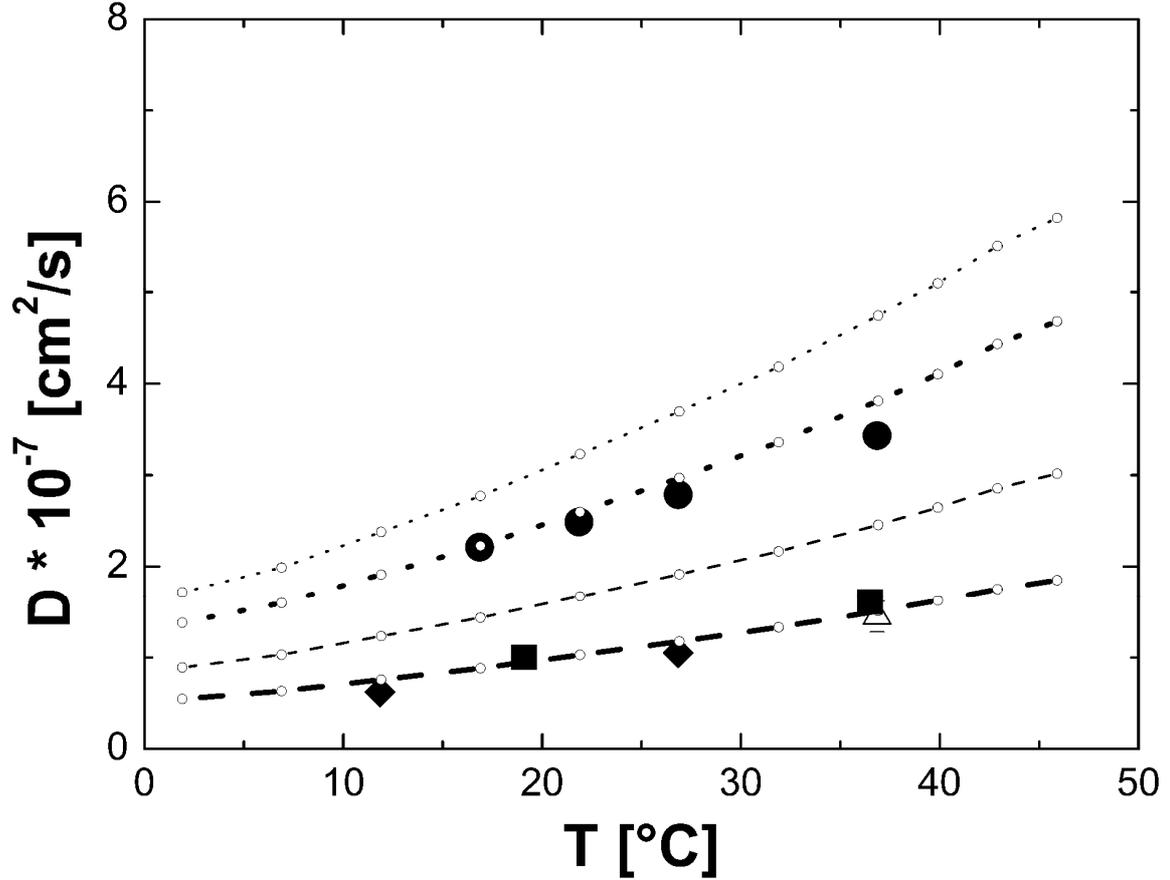

**Figure 3:** Diffusion coefficient $D$ of Hb in RBC as a function of temperature. The diffusion coefficients were measured with QENS on IRIS (circles), on IN16 (diamonds), on IN10 (squares) and with neutron spin-echo spectroscopy on IN15 (triangle) (Doster and Longeville 2007). $D_0$ was measured with dynamic light scattering and scaled data points are given as small empty circles. The thin dotted and the thin dashed line show the theoretical values of short-time $D_S^S$ and long-time self-diffusion $D_S^L$ of Hb at a volume fraction of $\phi=0.25$ with hydrodynamic interactions ($D_S^S = 0.56 \cdot D_0$, $D_S^L = 0.28 \cdot D_0$) (Tokuyama and Oppenheim 1994). The thick dotted and the thick dashed line represent the theoretical values for short-time and long-time self-diffusion of Hb assuming that a hydration layer of 0.23g $H_2O$/g protein is bound to the surface of Hb. Experimental data agree well with the theoretical considerations when the bound hydration water layer is taken into account ($D_S^S = 0.45 \cdot D_0$, $D_S^L = 0.18 \cdot D_0$) (Tokuyama and Oppenheim 1994).

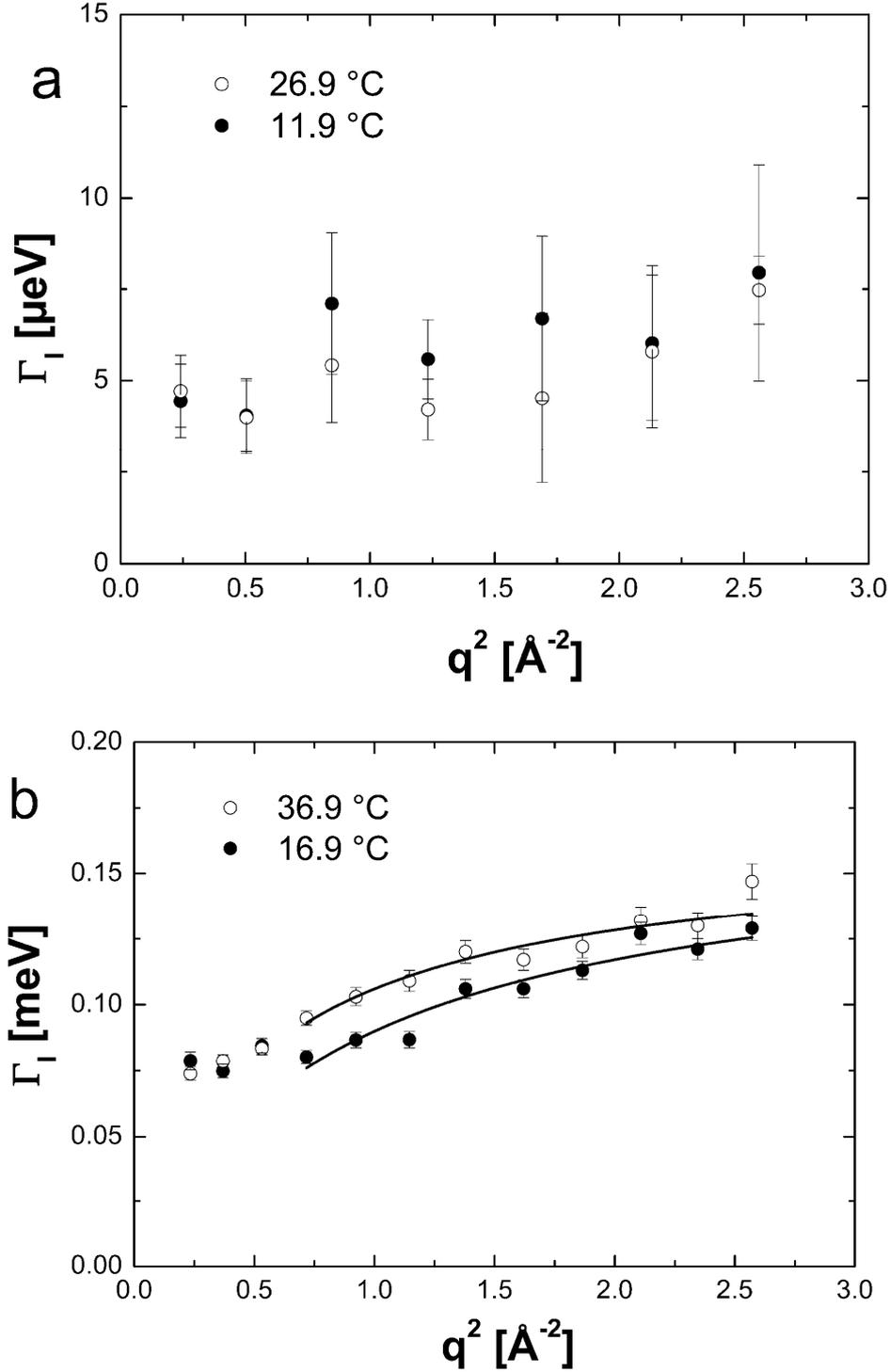

**Figure 4:** Half-widths at half-maximum of the half-widths at half-maximum $\Gamma_I(q)$ of internal Hb dynamics as a function of $q^2$. Data in (a) was measured on IN16 and (b) on IRIS. The scattering vector dependence of the line-widths contains information of the observed motions in Hb. Solid lines in (b) are fits with a jump-diffusion model in the $q^2$-range from 0.72 to 2.57 Å$^{-2}$

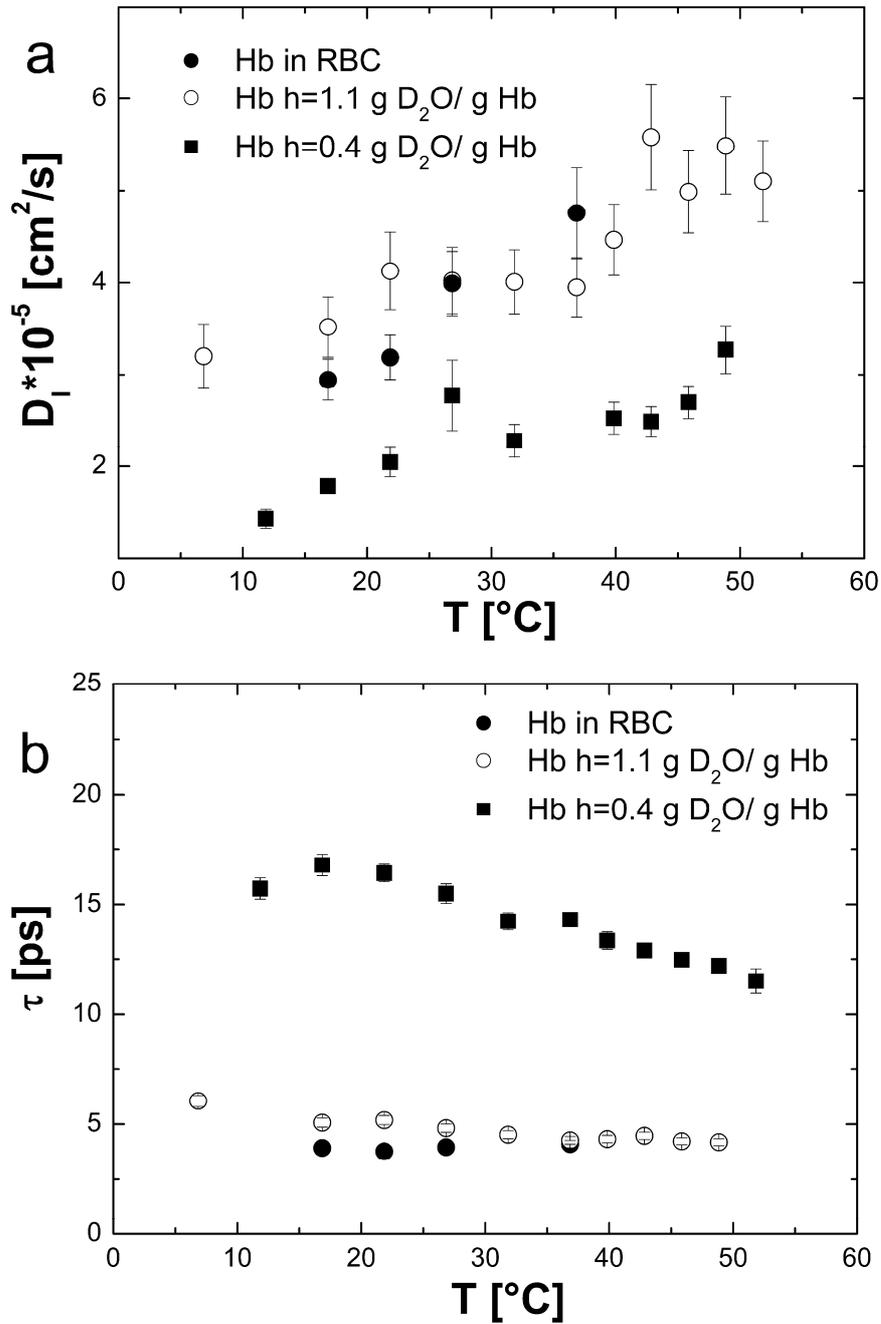

**Figure 5:** (a) Diffusion coefficients of internal motions in Hb as a function of temperature. Results from fully hydrated Hb powder, solution and Hb in RBC at different hydration levels are compared. (b) Residence times of internal jump-diffusion as a function of temperature of the different samples. Hb in RBC was measured on IRIS at ISIS (energy resolution 17μeV), concentrated Hb solution on TOFTOF at FRM-II (resolution 100μeV) and hydrated Hb powder on FOCUS at PSI (resolution 50μeV).

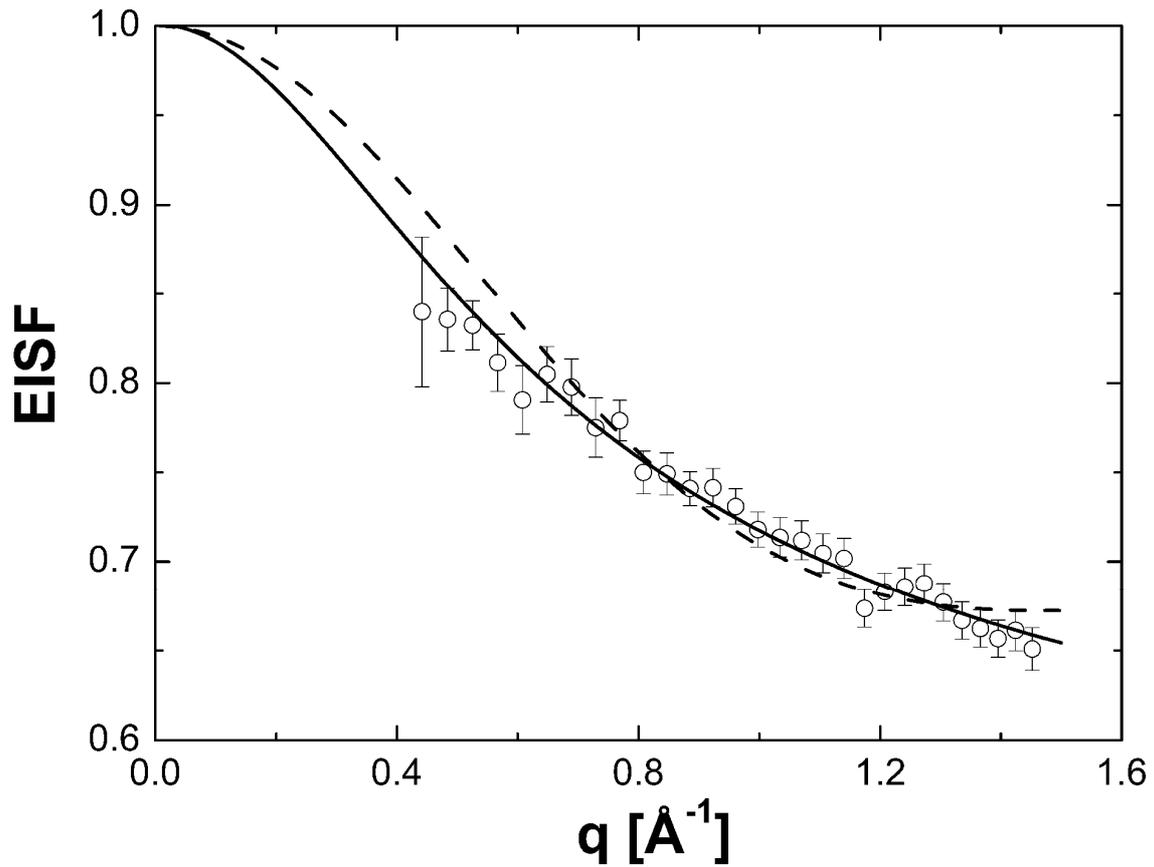

**Figure 6:** Elastic Incoherent Structure Factor of Hb in RBC measured on IRIS at the temperature 26.9 °C. The dashes line is a fit with the model for diffusion in a sphere. The solid line is a fit with the model for diffusion in a sphere with a Gaussian distribution of sphere radii.

## Short title for page headings

Macromolecular Dynamics in Red Blood Cells